\documentclass[twocolumn]{aastex63}
\usepackage{lineno}
\usepackage{inputenc}
\usepackage{booktabs}
\usepackage{latexsym}
\usepackage{graphicx}
\usepackage{amssymb}
\usepackage{longtable}
\usepackage{epsf}
\usepackage{lineno}
\usepackage{float}
\usepackage{comment}
\usepackage[]{natbib}
\usepackage{placeins}
\usepackage{multirow}
\usepackage{amsmath} 
\usepackage{hyperref}
\usepackage{booktabs}

\graphicspath{{./figures/}}

\begin{document}
\title{Ionized Gas Outflows in the Galaxy And Mass Assembly (GAMA) Survey: Signatures of AGN Feedback in Low-Mass Galaxies}

\author[0000-0002-4587-1905]{Sheyda Salehirad}
\affiliation{eXtreme Gravity Institute, Department of Physics, Montana State University, Bozeman, MT 59717, USA }
\author[0000-0001-7158-614X]{Amy E.\ Reines}
\affiliation{eXtreme Gravity Institute, Department of Physics, Montana State University, Bozeman, MT 59717, USA }
\author[0000-0001-8440-3613]{Mallory Molina}
\affiliation{Department of Physics \& Astronomy, University of Utah, James Fletcher Building, 115 1400 E, Salt Lake City, UT 84112, USA} \affiliation{Department of Physics \& Astronomy, Vanderbilt University, Nashville, TN 37235, USA}

%%%%%%%%%%%%%%%%%%%%%%%%%%%%%%%%%%%%%%%%%%%%%%%%%%%%%%%%%%%%%%%%%%%%%%%%%%%%%%%%%%%%%%%%%%%%%%%%%%%%%%%%%%%%%%%%%%%%%%%%%%%%%%%%%%%%%%%%%%%%%%%%%%%%%%%%%%%%%%%%%%%%%%%%%%%%%%
%%%%%%%%%%%%%%%%%%%%%%%%%%%%%%%%%%%%%%%%%%%%%%%%%%%%%%%%%%%%%%%%%%%%%%%%%%%%%%%%%%%%%%%%%%%%%%%%%%%%%%%%%%%%%%%%%%%%%%%%%%%%%%%%%%%%%%%%%%%%%%%%%%%%%%%%%%%%%%%%%%%%%%%%%%%%%%
\begin{abstract}
We present a sample of 398 galaxies with ionized gas outflow signatures in their spectra from the Galaxy and Mass Assembly (GAMA) Survey Data Release 4, including 45 low-mass galaxies with stellar masses $M_*<10^{10}$ $M_\odot$.
We assemble our sample by systematically searching for the presence of a second velocity component in the [\ion{O}{3}]$\lambda\lambda 4959, 5007$ doublet emission line in 39,612 galaxies with redshifts $z<0.3$.  
{The host galaxies are classified using the BPT diagram, with $\sim$89\% identified as AGNs and composites and 11\% as star-forming (SF) galaxies.}
The outflows are typically faster in AGNs with a median velocity of 936 km s$^{-1}$ compared to 655 km s$^{-1}$ in the SF objects. 
Of particular interest are the 45 galaxies in the low-mass range, of which a third are classified as AGNs/composites. 
The outflows from the low-mass AGNs are also faster and more blueshifted compared to those in the low-mass SF galaxies. This indicates that black hole outflows can affect host galaxies in the low-mass range and that AGN feedback in galaxies with $M_*<10^{10}$ $ M_\odot$ should be considered in galaxy evolution models.
\end{abstract}

%%%%%%%%%%%%%%%%%%%%%%%%%%%%%%%%%%%%%%%%%%%%%%%%%%%%%%%%%%%%%%%%%%%%%%%%%%%%%%%%%%%%%%%%%%%%%%%%%%%%%%%%%%%%%%%%%%%%%%%%%%%%%%%%%%%%%%%%%%%%%%%%%%%%%%%%%%%%%%%%%%%%%%%%%%%%%%
%%%%%%%%%%%%%%%%%%%%%%%%%%%%%%%%%%%%%%%%%%%%%%%%%%%%%%%%%%%%%%%%%%%%%%%%%%%%%%%%%%%%%%%%%%%%%%%%%%%%%%%%%%%%%%%%%%%%%%%%%%%%%%%%%%%%%%%%%%%%%%%%%%%%%%%%%%%%%%%%%%%%%%%%%%%%%%
\section{Introduction}\label{sec:intro}
Supermassive black holes (BHs) with masses $M_{\rm BH} \sim 10^{6-10} M_\odot$ are found in the nuclei of almost all massive galaxies \citep{Kormendy:1995,Kormendy:2013} and these BHs have primarily grown over cosmic time through merger-driven accretion. Observations show a tight correlation between BH mass and both galaxy bulge mass and stellar velocity dispersion \citep[e.g.,][]{Gebhardt:2000,Ferrarese:2000,Tremaine:2002,Marconi:2003,Gultekin:2009,McConnell:2013}, suggesting a connection between the evolution of these BHs and their host galaxies. Theoretical models indicate that this co-evolution can be regulated by feedback from an active galactic nucleus \citep[AGN;][]{Silk:1998,Churazov:2005,Somerville:2008,Vogelsberger:2014,Schaye:2015}. AGN feedback via radiation and outflows can impact the interstellar medium (ISM) in a galaxy and heat or eject the gas, ultimately inhibiting star formation and BH growth.

Without incorporating AGN feedback in galaxy models, key observational properties of galaxies such as the sharp cut-off at the end of the galaxy luminosity function can not be reproduced \citep[e.g.,][]{Bower:2006,Baldry:2012}. AGN feedback explains the red color of spheroidal galaxies, the lack of super bright and very massive galaxies, and the X-ray temperature-luminosity relationship \citep[e.g.,][]{Markevitch:1998,Benson:2003,Croton:2006,Menci:2006,McCarthy:2010}.  

Evidence for outflows has been found in both nearby galaxies \citep{Veilleux:2005,Rupke:2011,Mullaney:2013,Balmaverde:2016}, and high-redshift objects \citep{Alexander:2010,Carniani:2015}, and in ionized, atomic, and molecular phases of gas \citep{Feruglio:2010,Cicone:2012,Cicone:2014,Cazzoli:2014,Morganti:2017,Veilleux:2020,Fluetsch:2021}.  
Outflows produced by AGNs are usually thought to reduce or quench star formation 
\citep[e.g.,][]{Springel:2005,Hopkins:2006,Pereira-Santaella:2018,Ellison:2021}, however, in some cases, it can increase star formation activity \citep{Silk:2013,Zubovas:2013,Cresci:2015,Schutte:2022}.
We know the radiation fields and jets from accretion disks can launch outflows in AGNs, while star formation (e.g., stellar winds and supernovae) can also produce them \citep[see,][for reviews]{Heckman:2017,Rupke:2018,Wylezalek:2018,Veilleux:2020}. However, many of the exact driving mechanisms are still elusive.  It is often unknown if the same mechanisms drive different phases of outflows and if they have similar spatial distributions. We also usually do not know the exact morphology of the outflows (shell-like or conical). 

It has long been believed that stellar feedback is the main source of feedback in low-mass galaxies \citep[e.g.,][]{Martin-Navarro:2018}. However, the rising observational evidence of AGNs \citep{Reines:2013,Moran:2014,Sartori:2015,Baldassare:2016,Molinafex:2021,Salehirad:2022, Reines:2022} as well as recent findings of AGN-driven outflows in this mass range \citep{Penny:2018,Manzano:2019,Liu:2020,Aravindan:2023} suggest the notable impact of AGN feedback in these objects. Theoretical models have proposed contrasting results regarding AGN feedback in low-mass galaxies. Some studies found that AGN feedback quenches star formation \citep{Dashyan:2018,Barai:2019}, while others predict a negligible effect from AGN feedback contrary to stellar feedback \citep{Trebitsch:2018}. Additionally, some simulations indicate that AGN feedback increases the outflow energetics in these galaxies \citep{Koudmani:2019,Koudmani:2021}.

{The existence of broad-line features in the [\ion{O}{3}]$\lambda\lambda4959, 5007$ doublet line profile is a well-established tracer of ionized gas outflows \citep{Stockton:1976,Heckman:1981}. Given the forbidden transition of this line, [\ion{O}{3}] can not be produced in the dense broad line region (BLR) around an AGN and it can be used to investigate ionized gas dynamics in the narrow line region (NLR). Strong velocity gradients in the NLR associated with outflows can be observed as broadening or shifting of the [\ion{O}{3}] line that exceeds normal galaxy dynamics and can extend to kiloparsecs \citep[e.g.,][]{Pogge:1989}.}

In this paper, we search for ionized gas outflows in a large sample of galaxies with spectra in the Galaxy and Mass Assembly (GAMA) Survey Data Release 4 \citep[DR4;][]{Liske:2015,Driver:2022}. Our primary focus is on identifying outflows in low-mass galaxies with stellar masses $M_\star < 10^{10} M_\odot$.
The GAMA survey covers equatorial and southern sky regions and has minor overlaps with galaxies in the Sloan Digital Sky Survey (SDSS). GAMA is also two magnitudes deeper than the SDSS. While extensive research has been conducted on galaxies exhibiting outflow signatures in SDSS \citep[e.g,][]{Mullaney:2013,Matzko:2022},  the GAMA survey remains relatively unexplored. {Therefore, our objective is to discover new galaxies displaying ionized outflow signatures among both massive and low-mass objects, including AGNs and SF galaxies. We aim to analyze the outflow properties of these objects, specifically their offset and outflow velocities.} 

Section \ref{sec:data} details our data and sample selection process. Sections \ref{sec:analysis} and \ref{sec:outflow_candidates} present the analysis and results. The summary and conclusions can be found in Section \ref{sec:discussion}. We assume a $\Lambda$CDM cosmology with $\Omega_m=0.3$, $\Omega_\Lambda=0.7$ and $H_0 = 70$ km s$^{-1}$ Mpc$^{-1}$.

%%%%%%%%%%%%%%%%%%%%%%%%%%%%%%%%%%%%%%%%%%%%%%%%%%%%%%%%%%%%%%%%%%%%%%%%%%%%%%%%%%%%%%%%%%%%%%%%%%%%%%%%%%%%%%%%%%%%%%%%%%%%%%%%%%%%%%%%%%%%%%%%%%%%%%%%%%%%%%%%%%%%%%%%%%%%%%
%%%%%%%%%%%%%%%%%%%%%%%%%%%%%%%%%%%%%%%%%%%%%%%%%%%%%%%%%%%%%%%%%%%%%%%%%%%%%%%%%%%%%%%%%%%%%%%%%%%%%%%%%%%%%%%%%%%%%%%%%%%%%%%%%%%%%%%%%%%%%%%%%%%%%%%%%%%%%%%%%%%%%%%%%%%%%%
\section{Data and Parent Sample}\label{sec:data}

\subsection{The GAMA Survey}\label{sec:GAMA_survey}
We use the publicly available data from the GAMA Survey DR4 \citep{Liske:2015,Driver:2022} to conduct our study. The GAMA survey comprises optical spectroscopic observations taken with the AAOmega multi-object spectrograph \citep{Saunders:2004, Smith:2004, Sharp:2006} on the 3.9 m Anglo-Australian Telescope. The wavelength range of the dual-beam set-up of the spectrograph is 3730--8850 \AA, the spectral resolution of the blue and red arms are 3.5 and 5.3 \AA, respectively, and the spectroscopic fibers are 2" in diameter. The survey covers 2 southern regions (G02 and G23), with respective areas of 56 and 51 deg$^2$, and three 60 deg$^{2}$ equatorial regions (G09, G12 and G15). The magnitude limit for the main survey of galaxies in the equatorial and G02 regions is each ${\rm r}<19.8$ mag and the limiting magnitude of the G23 region is ${\rm i}<19.8$ mag \citep{Baldry:2018,Driver:2022}.

\subsection{Parent Sample of Galaxies}\label{sec:sample_selection}
We define our parent sample using the spectra provided in the \texttt{AATSpecAll v27} table in the \texttt{SpecCat} data management unit \citep[DMU;][]{Liske:2015}  and following the \citet{Salehirad:2022} methodology, {as described here}. To ensure high-quality data, {we select spectra with \texttt{COMMENTS\_FLAG} $=0$, which excludes unreliable detections such as those with fringing and bad splicing. We choose spectra with normalized redshift values of \texttt{NQ}$>2$, as suggested by GAMA, corresponding to a minimum 90\% probability that the best redshift estimate is accurate. If multiple observations are available for a galaxy, we select the spectrum with the best redshift using the column value \texttt{IS\_SBEST} $=1$.  The redshift is determined using cross-correlation of spectra and stellar templates, with the best-estimated redshift adopted from the highest cross-correlation peak, normalized by a root mean square value. The confidence in this redshift estimate is assessed by comparing the height of the highest correlation peak with those of the next three best redshift estimates \citep[][and the references therein]{Liske:2015}.} Finally, we only include objects with redshifts ${\rm z}<0.3$ to ensure the lines of interest such as the [\ion{S}{2}] doublet are detected.

We then apply signal-to-noise ratio (S/N) cuts similar to \citet{Reines:2013}, using the emission-line fluxes and equivalent width (EW) measurements given in the \texttt{GaussFitSimple v05} table from the \texttt{SpecLineSFR} DMU \citep{Gordon:2017}. We select galaxies that have ${\rm S/N}\geq3$ and ${\rm EW}>1$~\AA\ for the H$\alpha$, [\ion{O}{3}]~$\lambda5007$ and [\ion{N}{2}]~$\lambda6583$ lines. Given that the H$\beta$ line is generally a weaker line compared to the H$\alpha$ line, we select those with ${\rm S/N}\geq2$.

The stellar masses are stored in the \texttt{StellarMasses} DMU \citep{Taylor:2011} and in various tables depending on the sky regions. We include all the galaxies with available stellar mass estimates and impose a mass cut of ${\rm M_\star}>10^{5}$ M$_\odot$ to avoid possible star detections. Stellar masses for galaxies in the G23 region are only given in the \texttt{StellarMassesGKV v24} table \citep{Driver:2022} which also contains stellar masses for the equatorial galaxies. Here, the masses are derived using all band photometry from the Kilo-Degree Survey \citep[KiDS;][]{Kuijken:2019} and the Visible and Infrared Survey Telescope for Astronomy Kilo-degree Infrared Galaxy Public Survey \citep[VIKING;][]{Edge:2013}. Tables \texttt{StellarMassesG02SDSS v24} and \texttt{StellarMassesG02CFHTLS v24} provide stellar masses for the G02 galaxies. The values in these tables are derived by multi-band spectral energy distribution (SED) fitting to SDSS and Canada-France-Hawaii Telescope Lensing \citep[CFHTLenS;][]{Heymans:2012} photometry, respectively. We utilize the stellar masses given in the \texttt{StellarMassesG02CFHTLS v24} table but use the \texttt{StellarMassesG02SDSS v24} table to exclude the galaxies with masses that are different by 0.3 dex in both tables. Our final sample consists of 39,612 galaxies.

%%%%%%%%%%%%%%%%%%%%%%%%%%%%%%%%%%%%%%%%%%%%%%%%%%%%%%%%%%%%%%%%%%%%%%%%%%%%%%%%%%%%%%%%%%%%%%%%%%%%%%%%%%%%%%%%%%%%%%%%%%%%%%%%%%%%%%%%%%%%%%%%%%%%%%%%%%%%%%%%%%%%%%%%%%%%%%
%%%%%%%%%%%%%%%%%%%%%%%%%%%%%%%%%%%%%%%%%%%%%%%%%%%%%%%%%%%%%%%%%%%%%%%%%%%%%%%%%%%%%%%%%%%%%%%%%%%%%%%%%%%%%%%%%%%%%%%%%%%%%%%%%%%%%%%%%%%%%%%%%%%%%%%%%%%%%%%%%%%%%%%%%%%%%%

\section{{Spectral Analysis}}\label{sec:analysis} 

In this paper, we systematically search for double-component features in the [\ion{O}{3}]$\lambda\lambda 4959,5007$ doublet lines that may signify outflows. 
Complex emission line profiles exhibiting asymmetries, shoulders, and/or double peaks can indicate the presence of two or more gaseous components with distinct kinematics along the line of sight. However, non-Gaussian line structure can also result from beam-spearing of the velocity gradient due to relatively coarse spatial resolution \citep{2015:Garcia-Lorenzo}.

In previous studies, the broad blueshifted wings in the [\ion{O}{3}] line profile have been interpreted as outflows \citep[e.g.,][]{Mullaney:2013,Zakamska:2014,Harrison:2016,Geach:2018,Guolo-Pereira:2021}, constituting around 40\% of the overall flux of the [\ion{O}{3}] line \citep[e.g.,][]{Concas:2017}. Fewer studies have focused on broad redshifted lines in the [\ion{O}{3}] line profile, but they could also be attributed to outflows depending on the inclination of the galaxy with respect to the line of sight \citep{Crenshaw:2010,Bae:2016}. In cases of bipolar outflows and depending on the orientation, we can also observe symmetric broad lines \citep{Harrison:2012}. {Therefore, to accumulate a more comprehensive outflow sample, we do not limit our search to blueshifted broad wings and freely select the center of the second component. }

While our parent sample is determined using the emission-line flux measurements provided in the GAMA survey, we write our custom code to analyze the spectra and fit the emission lines. We subsequently visually inspect each flagged galaxy spectrum and exclude the unreliable ones, such as those with bad fits (fits to the noise in the spectrum), those with missing pixel values within the emission lines, and those affected by bad splicing or fringing. {In this section, we describe our method for fitting the continuum, selecting outflow candidates, evaluating outflow velocities, and modeling other relevant emission lines.}

%%%%%%%%%%%%%%%%%%%%%%%%%%%%%%%%%%%%%%%%%%%%%%%%%%%%%%%%%%%%%%%%%%%%%%%%%%%%%%%%%%%%%%%%
%%%%%%%%%%%%%%%%%%%%%%%%%%%%%%%%%%%%%%%%%%%%%%%%%%%%%%%%%%%%%%%%%%%%%%%%%%%%%%%%%%%%%%%%

\subsection{Stellar Continuum Subtraction} \label{sec:continuum}    

\begin{figure*}[htbp]
\centering
\includegraphics[width=\textwidth]{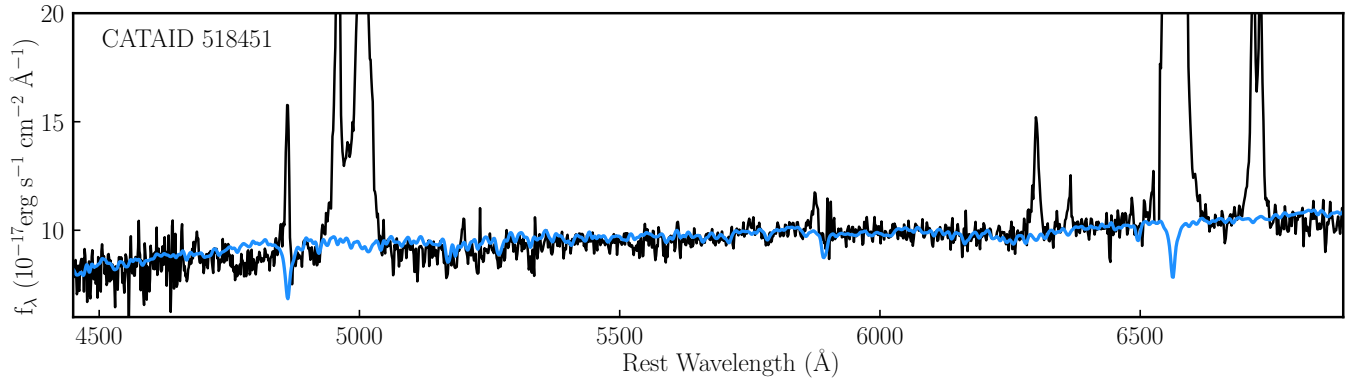}
\caption{{An example of stellar continuum fit (AGN with CATAID 518451). Here the redshift-corrected spectrum is shown in black and the best-fitted stellar continuum model is in blue. See Section \ref{sec:continuum} for details.} }
\label{fig:spec_line_example}
\end{figure*}

We use the  Penalized Pixel-Fitting method \cite[\texttt{pPXF};][]{Cappellari:2017} to fit the stellar continuum for each galaxy. We utilize the stellar population synthesis models from  \citet{Bruzual:2003} as our stellar continuum templates in 3 metallicities ${\rm (Z=0.008, 0.02, 0.05)}$ and 10 ages (t$=$0.005, 0.025, 0.1, 0.29, 0.64, 0.9, 1.4, 2.5, 5, and 11 Gyr). Initially, we model the spectra with a combination of single-metallicity templates of various ages, modified by a low-order multiplicative polynomial to account for dust reddening, and select the model metallicity with the smallest $\chi^2$ value. This method produces reliable continuum fits for the majority of the galaxies in our sample. However, the initial continuum fit in some cases yields unrealistically large velocity dispersion values given by \texttt{pPXF}, which is evident in the extreme broadening of the absorption line fits. For these objects, we redo the fits by changing the order of the multiplicative polynomials, adding additive polynomials as they can minimize the template mismatch by changing the strength of the absorption lines \citep{Cappellari:2017}, or changing the wavelength range of the spectrum to cut the noisy ends. Since our primary goal is to measure the emission lines, we attempt good fits to the stellar continua but do not fully explore the parameter space. An example of a fitted galaxy spectrum in our sample is shown in Figure \ref{fig:spec_line_example}.

We also find a handful of AGN-dominated spectra among the flagged galaxies in which the stellar templates do not provide an optimal fit to the continuum. However, since we include a linear component in the fit of the emission lines of interest (see below), the [\ion{O}{3}] doublet lines are fitted sufficiently in the end.

%%%%%%%%%%%%%%%%%%%%%%%%%%%%%%%%%%%%%%%%%%%%%%%%%%%%%%%%%%%%%%%%%%%%%%%%%%%%%%%%%%%%%%%%
%%%%%%%%%%%%%%%%%%%%%%%%%%%%%%%%%%%%%%%%%%%%%%%%%%%%%%%%%%%%%%%%%%%%%%%%%%%%%%%%%%%%%%%%
\subsection{Fitting the [\ion{O}{3}] lines} \label{sec:analysis_o3_lines}

\begin{figure}[htbp]
\centering
\includegraphics[width=0.48\textwidth]{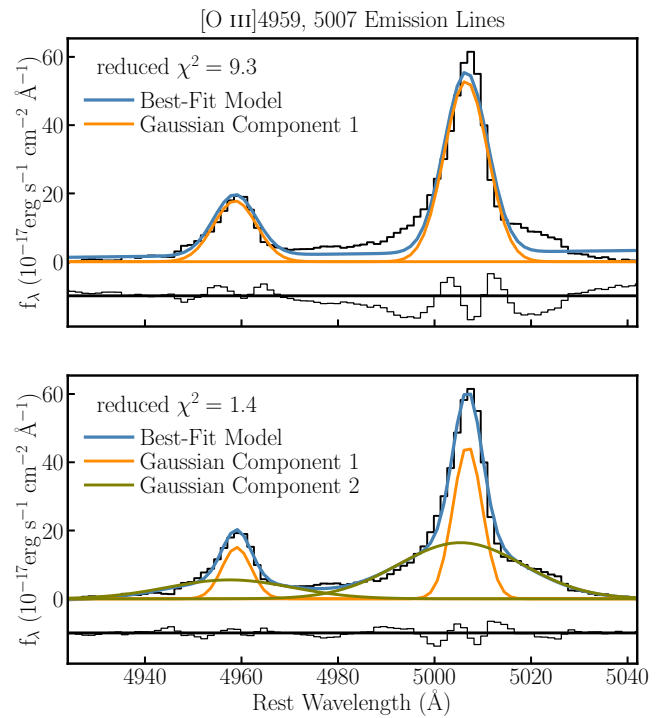}
\caption{{The [\ion{O}{3}] doublet emission line plotted with two different fitting models. The top panel shows a single Gaussian model fitted to each line, while the bottom panel displays two Gaussian fits. 
The black line represents the observed emission line. The blue line indicates the best-fitting model, which consists of Gaussian and linear components. The orange line shows the narrow (systemic) component, while the olive line depicts the broad (outflow) component. Residuals are displayed in black with a vertical offset. Adding a second component into the [\ion{O}{3}] fit for this galaxy significantly improves the final model. For more information, see Section \ref{sec:analysis_o3_lines}. }} 
\label{fig:o3fit}
\end{figure}

We use the \texttt{LMFIT} package in python \citep{lmfit} to fit chunks of spectra around the [\ion{O}{3}]$\lambda\lambda4959,5007$ doublet. We incorporate Gaussian models to fit the emission lines and a linear model which accounts for the continuum fit residuals.

\begin{figure*}[t]
\centering
\includegraphics[width=\textwidth]{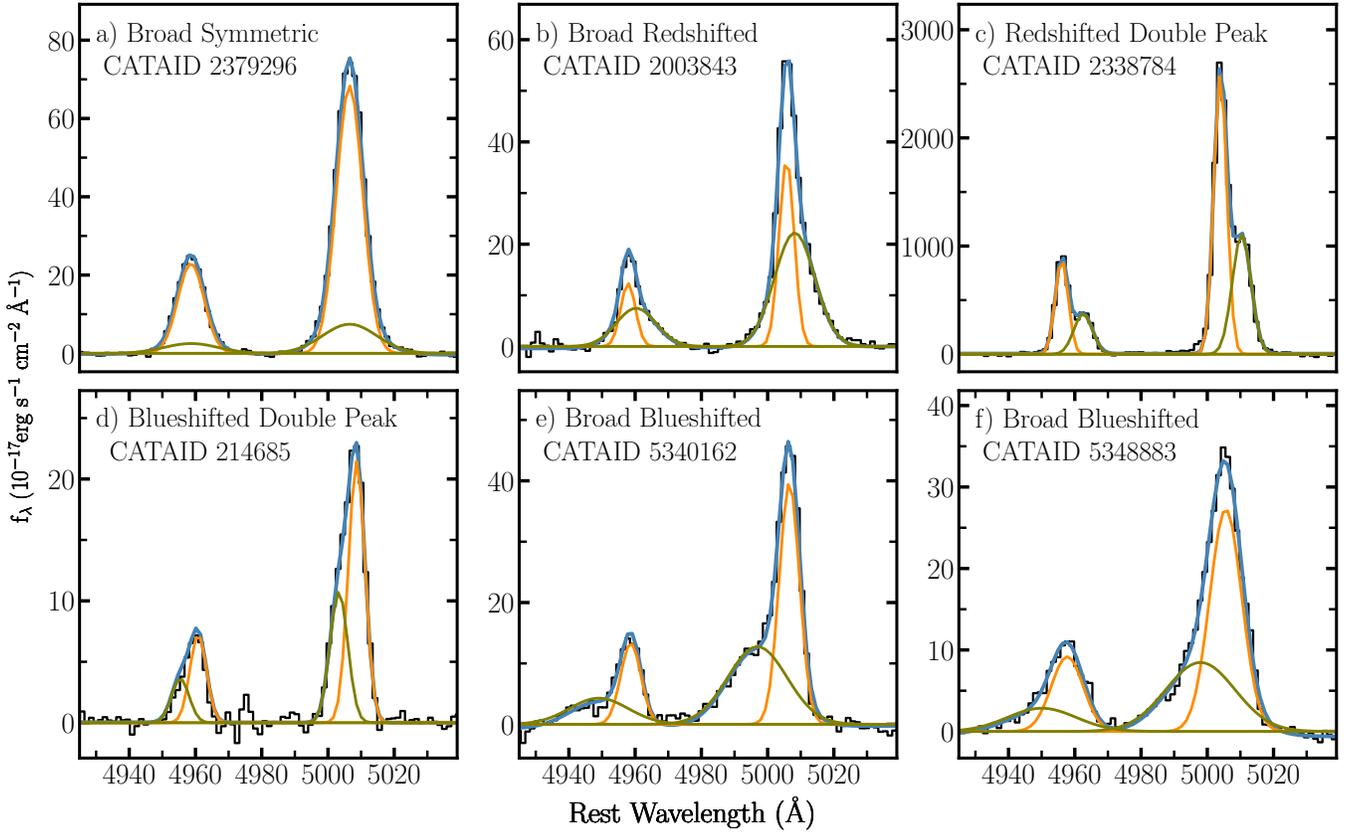}
\caption{{Examples of the [\ion{O}{3}]$\lambda\lambda 4959, 5007$ doublet line profiles from our outflow sample, fitted using two-Gaussian models. The color scheme matches that of Figure \ref{fig:o3fit}. 
Panels (a) and (b) show cases with broad symmetric and broad redshifted outflow lines, respectively. Panels (c) and (e) present lines exhibiting two peaks, with components that have similar widths.  
Panels (e) and (f) show broad blueshifted lines.   In panel (e), the overall line profile indicates a blueshifted bump, while the components are blended in panel (f). } For further details, refer to Section \ref{sec:analysis_o3_lines}.} 
\label{fig:red_dp_lines_examples}
\end{figure*}

We fit the [\ion{O}{3}]$\lambda\lambda4959,5007$ doublet lines simultaneously with both one- and two-Gaussian models. In the one-Gaussian model, the lines in the doublet are restricted to have equal velocity widths, fixed laboratory separation, and a flux ratio of [\ion{O}{3}]$\lambda5007$/[\ion{O}{3}]$\lambda4959$ $=$ 3. In the two-Gaussian model, we allow the first and second components of the [\ion{O}{3}]$\lambda5007$ line to change freely, however, the [\ion{O}{3}]$\lambda4959$ line components have the same velocity widths, fixed separation, and fixed flux ratio to the respective [\ion{O}{3}]$\lambda5007$ line components. The two-component model is adopted if the reduced $\chi^2$ is lowered by at least 20\% relative to the one-component model, and the second components are broader than the first components. We only select lines with widths that are at least equal to the instrumental spectral resolution ($\rm v_{FWHM} \gtrsim 200$ km s$^{-1}$), those that have a flux $\rm S/ \rm N \geq3$ for each of the Gaussian components, and second components with line peaks that are at least 3$\sigma$ above the root mean square (rms) noise level of the continuum chunk. Figure \ref{fig:o3fit} shows an example of modeled emission lines.

{We identified 439 galaxies with outflow signatures in the [\ion{O}{3}]$\lambda\lambda 4959,5007$ lines.} After visually inspecting the fits of all the flagged candidates, we remove 41 objects with unreliable fits. These cases consist of objects with spectra affected by fringing and bad spicing and those with lines fit to the continuum noise. {Our final sample consists of 398 galaxies. }

We find outflow components that are symmetric with respect to the narrow (systemic) component, as well as those that are blueshifted or redshifted.
{Some of the identified second components contribute to as little as 5\% of the total flux of the [\ion{O}{3}] line, potentially representing weaker outflows or non-Gaussian profiles. Despite their weaker nature, these second components meet the necessary criteria, and hence, we include them in our final sample.} In 9\% of the outflow candidates, we observe either two distinct peaks in the overall [\ion{O}{3}] line profile or less pronounced double peaks, but lines with similar widths. {These lines can be what some studies call double-peaked lines, where they are usually associated with disk rotation of the NLR around a single BH, biconical outflows, or distinct NLRs in merging AGNs \cite[e.g.,][]{Shen:2011}. Indeed, these lines have been found in single AGNs \citep[e.g.,][]{Muller:2015,Nevin:2018,Bizyaev:2022} as well as dual AGNs \citep{Rosario:2011,Fu:2023}.}
Examples of [\ion{O}{3}] doublet emission-line profiles are shown in Figure \ref{fig:red_dp_lines_examples}.

\subsection{Outflow Velocity} \label{sec:analysis_outflow_vel}

\begin{figure}[h]
\centering
\includegraphics[width=0.47\textwidth]{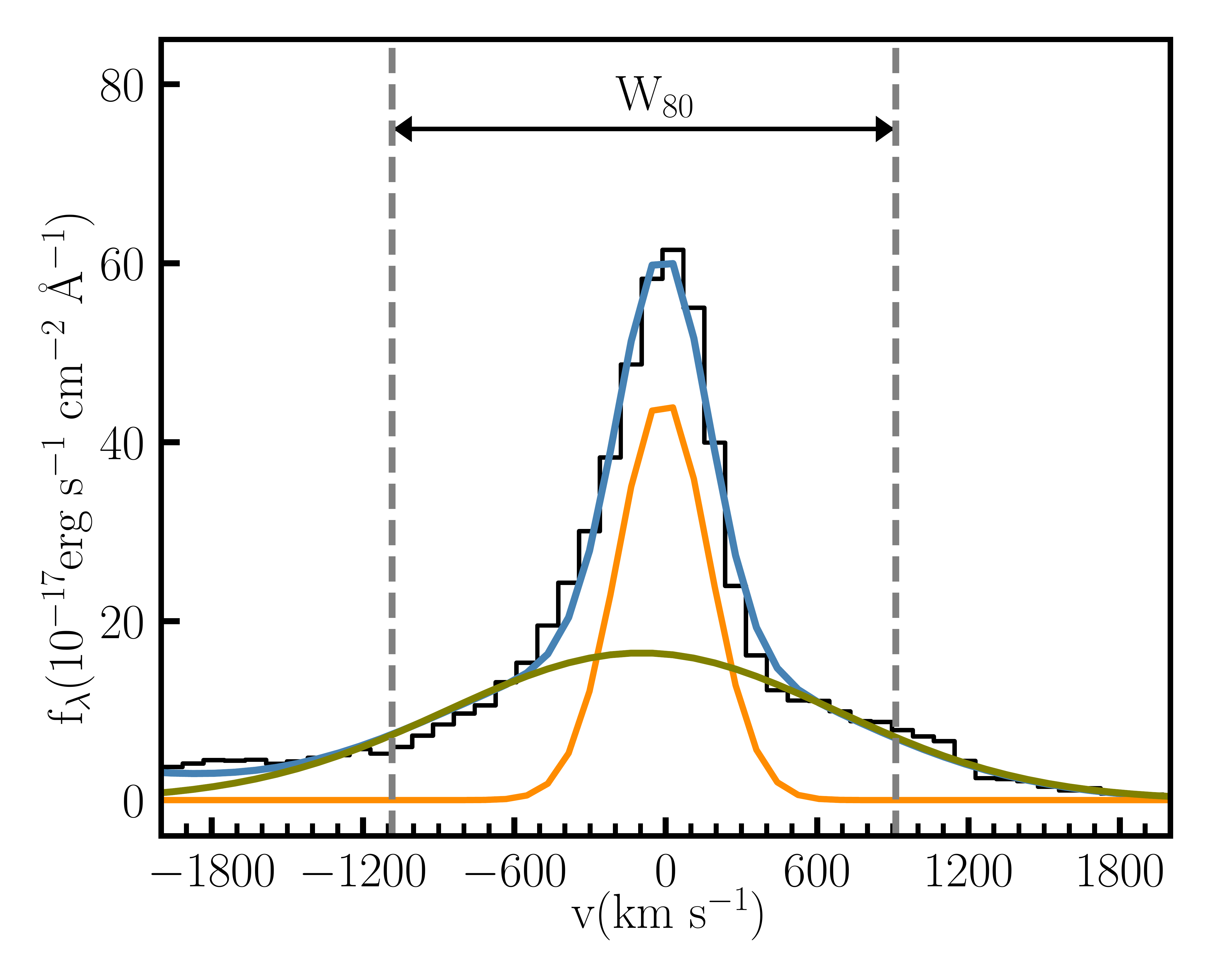}
\caption{{ This plot illustrates an example of W$_{80}\sim$ FWHM (dashed gray lines) for the outflow component (green line) from the [\ion{O}{3}]$\lambda5007$ emission, indicating outflow velocity.
See Section \ref{sec:analysis_outflow_vel} for details.} } 
\label{fig:w80}
\end{figure}

{In this work, we use W$_{80}$ to estimate outflow velocity, defined as the line width encompassing 80\% of the flux of the emission line \citep{Zakamska:2014}.
For a Gaussian profile, W$_{80}$ is related to the FWHM of the line and can be described as $\rm W_{80} = 1.09 \rm FWHM$ as shown in Figure \ref{fig:w80}. Some studies account for the offset velocity ($\rm v_o$) between the outflow and systemic emission-line components when evaluating the outflow velocity \citep[e.g.,][]{Manzano:2019}. However, the offset velocity is sensitive to dust extinction and inclination effects \citep[][]{Harrison:2014,Bae:2016}, while $\rm W_{80}$ is less affected by these factors and can better reflect typical bulk motions \citep{Harrison:2014}. Considering the sensitivity of $\rm v_o$ to extinction and the fact that our sample includes both redshifted and blueshifted outflows (see Section \ref{sec:analysis_o3_lines}), we adopt $\rm W_{80}$ as our measure of outflow velocity.}

\subsection{Other Emission Line Measurements} \label{sec:analysis_other_lines}

\begin{figure*}[t]
\centering
\includegraphics[width=\textwidth]{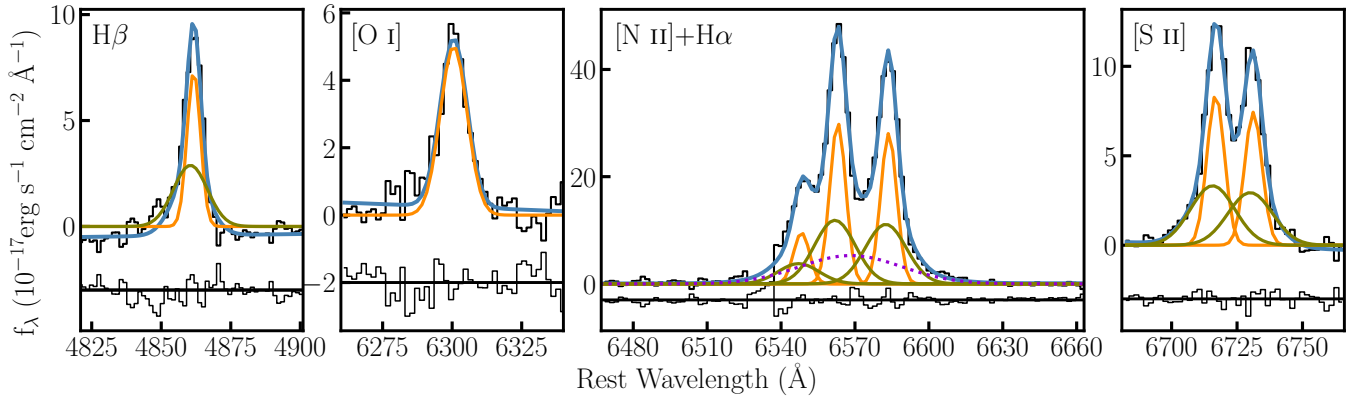}
\caption{{This figure shows chunks of emission-line spectra for H$\beta$, [\ion{O}{1}]$\lambda6003$, [\ion{N}{2}]+H$\alpha$ complex, and the [\ion{S}{2}]$\lambda\lambda 6716, 6731$ lines. The color scheme matches that of Figure \ref{fig:o3fit}. The dotted violet line represents the broad H$\alpha$ model. The residuals are plotted in black with a vertical offset. For more information, see Section \ref{sec:analysis_other_lines}.}} 
\label{fig:other_lines}
\end{figure*}

\begin{deluxetable*}{ccDcccccccc}
\tabletypesize{\footnotesize}
\tablecaption{Outflow Galaxies}
\tablehead{
\colhead{CATAID}&\colhead{RA}& \multicolumn2c{DEC}  & \colhead{$z$} &\colhead{log ($M_*/M_\odot$)} & \colhead{$g-r$} &\colhead{[\ion{N}{2}]/H$\alpha$}&\colhead{[\ion{S}{2}]/H$\alpha$}&\colhead{[\ion{O}{1}]/H$\alpha$}&\colhead{$v_0$} &\colhead{$v_{out}$} \\
\colhead{} & \colhead{{(degrees)}} & \multicolumn2c{{(degrees)}}& \colhead{ } & \colhead{ }  & \colhead{{(mag)}}  &\colhead{{Classification}}&\colhead{{Classification}}&\colhead{{Classification}}&\colhead{{(km s$^{-1}$)} }&\colhead{{(km s$^{-1}$)}}
}
\decimalcolnumbers
\startdata
1560359 & 30.28563 & -4.67478 & 0.21100 & 10.73 & 0.62 &     AGN &      Sy & \ldots & -408$\pm$50 &  1415$\pm$80 \\
1534204 & 30.75825 & -6.25522 & 0.13416 & 10.29 & 0.39 &      SF &      SF &     SF &   -3$\pm$22 &  684$\pm$104 \\
1557617 & 30.82937 & -4.83862 & 0.13701 &  8.91 & 0.04 &      SF &      Sy &     Sy &   73$\pm$44 &  866$\pm$145 \\
1485133 & 30.97987 & -9.31706 & 0.29487 & 11.04 & 0.72 &     AGN &      Sy &     Sy &    47$\pm$8 &  1349$\pm$35 \\
1555278 & 31.00350 & -4.98673 & 0.25509 & 10.34 & 0.41 &     AGN &      Sy &     Sy & -113$\pm$24 &   925$\pm$65 \\
1568106 & 31.09625 & -4.14728 & 0.21588 & 11.01 & 0.62 &     AGN &      Sy &     Sy & -317$\pm$50 &  1416$\pm$64 \\
2230554 & 32.08583 & -5.35770 & 0.20913 & 10.88 & 0.60 &     AGN &      Sy &     Sy & -151$\pm$40 & 1381$\pm$118 \\
2202335 & 32.14975 & -4.17996 & 0.05719 & 10.05 & 0.56 &   Comp. &      SF &     SF &    0$\pm$54 & 1145$\pm$210 \\
1307713 & 32.82988 & -5.73585 & 0.04248 & 10.28 & 0.61 &     AGN &      Sy &     Sy &  125$\pm$16 &   836$\pm$25 \\
2204417 & 32.98467 & -3.82524 & 0.09473 & 10.81 & 0.56 &   Comp. &      Sy &     Sy &  212$\pm$79 &  829$\pm$114 \\
\enddata
\tablecomments{{Properties of} the outflow candidates in this work. Columns 1-6 are obtained from GAMA DR4 and assume $h = 0.7$. Column 1: Unique ID of the GAMA object. Columns 2--3: The right ascension and declination (in degrees) of the spectrum (J2000). Column 4: Redshift. Columns 5--6: The log galaxy stellar mass in units of $M_\odot$ and $g-r$ color. All values are obtained from the \texttt{StellarMassesG02CFHTLS v24} and \texttt{StellarMassesGKV v24} tables \citep{Taylor:2011,Bellstedt:2020}. Columns 7--9: Classifications of the galaxy in the narrow-line diagnostic diagrams. {Columns 10--11: The offset and outflow velocities derived from the components of the [\ion{O}{3}]$\lambda5007$ line in km s$^{-1}$ with errors included.}
A three-dot ellipsis indicates no line is detected, or the emission lines do not meet our reliable detection criteria.
The entirety of Table~\ref{tab:gal_prop} is published in the electronic edition of {\it The Astrophysical Journal}. We show a portion here to give information on its form and content.
}
\label{tab:gal_prop}
\end{deluxetable*}

\begin{deluxetable*}{ccccccccccc}
\tabletypesize{\footnotesize}
%\setlength{\tabcolsep}{1.5pt}
%\renewcommand{\arraystretch}{1.}
%\tablewidth{0pt}
\tablecaption{Emission-line Fluxes}
\tablehead{
\colhead{CATAID}&\colhead{H$\beta_n$}&\colhead{H$\beta_b$}& \colhead{[\ion{O}{3}]$\lambda5007_n$}& \colhead{[\ion{O}{3}]$\lambda5007_b$}&\colhead{[\ion{O}{1}]$\lambda6300$} &\colhead{H$\alpha_n$}  &\colhead{H$\alpha_b$}&\colhead{[\ion{N}{2}]$\lambda6583$} & \colhead{[\ion{S}{2}]$\lambda6716$} & \colhead{[\ion{S}{2}]$\lambda6731$}}
\decimalcolnumbers
\startdata
1560359 &   62$\pm$6 &   \ldots & 234$\pm$12 &  183$\pm$17 &     \ldots &   205$\pm$6 &   132$\pm$18 &   213$\pm$5 &    76$\pm$4 &    64$\pm$4 \\
1534204 & 802$\pm$13 &   \ldots & 366$\pm$41 &  233$\pm$39 &   87$\pm$4 & 3248$\pm$13 &   304$\pm$31 & 1114$\pm$12 &   598$\pm$8 &   408$\pm$7 \\
1557617 &  204$\pm$3 &   \ldots & 1372$\pm$9 &   76$\pm$12 &   15$\pm$2 &   787$\pm$2 &       \ldots &    24$\pm$2 &    36$\pm$4 &    33$\pm$4 \\
1485133 &   23$\pm$6 & 66$\pm$6 &  525$\pm$7 &   329$\pm$8 &   15$\pm$1 &   242$\pm$4 &   158$\pm$13 &   167$\pm$4 &    56$\pm$4 &    54$\pm$3 \\
1555278 &   46$\pm$3 &   \ldots & 154$\pm$15 &  146$\pm$15 &   27$\pm$4 &   204$\pm$3 &   115$\pm$18 &   101$\pm$3 &    27$\pm$4 &    39$\pm$5 \\
1568106 &  91$\pm$13 &   \ldots & 344$\pm$50 &  481$\pm$54 &   65$\pm$8 &  284$\pm$17 &   475$\pm$40 &  328$\pm$15 &   166$\pm$9 &   138$\pm$9 \\
2230554 &   46$\pm$5 &   \ldots & 168$\pm$16 &  133$\pm$16 &   22$\pm$4 &   139$\pm$5 &   221$\pm$17 &   137$\pm$5 &    50$\pm$3 &    40$\pm$3 \\
2202335 &  378$\pm$8 &   \ldots & 161$\pm$17 &  148$\pm$23 &   61$\pm$5 &  1907$\pm$5 &   238$\pm$19 &  1001$\pm$5 &   319$\pm$5 &   231$\pm$5 \\
1307713 & 650$\pm$17 &   \ldots & 170$\pm$48 & 2069$\pm$65 & 282$\pm$16 & 2736$\pm$31 & 1165$\pm$125 & 3066$\pm$22 & 826$\pm$208 & 704$\pm$177 \\
2204417 & 466$\pm$32 &   \ldots & 327$\pm$90 & 418$\pm$104 &  92$\pm$13 & 1401$\pm$27 &   731$\pm$66 &  934$\pm$23 &  371$\pm$15 &  278$\pm$14 \\
\enddata
\tablecomments{Emission line fluxes for our outflow sample. Column 1: Unique ID of the GAMA object. Columns 2--11: The emission-line fluxes and{ their errors} in units of 10$^{-17}$ erg s$^{-1}$ cm$^{-2}$.
No extinction correction has been applied. The subscripts n and b refer to the narrow and broad components, respectively. We do not show the flux values of the [\ion{O}{3}]$\lambda 4959$ and the [\ion{N}{2}]$\lambda 4548$ lines, since their fluxes are fixed to be weaker by factors of 3. A three-dot ellipsis indicates no line is detected, or the emission lines do not meet our reliable detection criteria.
The entirety of Table \ref{tab:flux} is published in the electronic edition of {\it The Astrophysical Journal}. We show a portion here to give information on its form and content.  }
\label{tab:flux}
\end{deluxetable*}

We also fit Gaussian models to the [\ion{S}{2}]$\lambda\lambda 6716, 6731$, [\ion{N}{2}]+H$\alpha$, H$\beta$, and [\ion{O}{1}]$\lambda6003$ emission lines in our outflow sample following the methodology in \citet{Reines:2013} and references therein{, as described below}. The profiles of these lines do not typically match the profile of the [\ion{O}{3}] lines and therefore, we fit them independently. We use the derived emission line fluxes with $\rm S/N \geq 3$ to classify the outflow candidates based on their location on two-dimensional diagnostic diagrams (see Section \ref{sec:narrow_diag}). An example of our fits for these lines is shown in Figure \ref{fig:other_lines} and the host galaxy properties and the emission-line fluxes of the outflow candidates are listed in Tables \ref{tab:gal_prop} and \ref{tab:flux}, respectively.

We fit the [\ion{S}{2}] lines with one- and two-Gaussian components. For the single Gaussian model, the widths of the lines are assumed to be equal (in velocity space) while the laboratory separation between the lines is held fixed. In the two-component case, we restrict the relative widths, heights, and centers of the two components to be the same for both lines in the doublet. If the reduced $\chi^2$ of the two-component model is at least 20\% less and the width of the second component is larger than the first component, we select the two-Gaussian model for the [\ion{S}{2}] lines. This is the case in 30 of the outflow candidates.

Given that the profiles of the [\ion{N}{2}]$\lambda\lambda6548,6583$ and H$\alpha$ lines often match the [\ion{S}{2}] lines \citep{Filippenko:1988,Filippenko:1989,Ho:1997,Greene:2004}, we use the parameters from the [\ion{S}{2}] doublet models to fit the [\ion{N}{2}]+H$\alpha$ complex. For the one-Gaussian model, we assume the [\ion{N}{2}] lines have equal velocity widths to that of the [\ion{S}{2}] lines, their laboratory relative wavelength separation is fixed, and the flux ratio of [\ion{N}{2}]$\lambda$6583/[\ion{N}{2}]$\lambda6548$ $=$ 3. The width of the H$\alpha$ line is allowed to increase as much as 25\%. For the 30 objects with two-component fits to their [\ion{S}{2}] doublet lines, we scale the widths, centers, and heights of their [\ion{N}{2}] and H$\alpha$ components to that of the [\ion{S}{2}] lines. 

To search for broad H$\alpha$ emission that could signify dense gas orbiting a BH, we fit the lines with an additional broad H$\alpha$ component and select this model if the full-width at half maximum (FWHM) of the broad H$\alpha$ line is at least 500 km s$^{-1}$ after correcting for the instrumental resolution, and the reduced $\chi^2$ of the model with the addition of the broad H$\alpha$ component is at least 20\% less than the one without. We identify 206 outflow galaxies
with broad H$\alpha$ emission. The H$\beta$ line is fitted using the same method as the H$\alpha$ line, using the [\ion{S}{2}] profile as a template for the narrow line.

We also fit the [\ion{O}{1}]$\lambda6003$ line with one- and two-Gaussian models and  
select the two-component model if the width of the second component is larger than that of the first one, and the reduced $\chi^2$ is lowered by at least 20\%. 
Given that the [\ion{O}{1}] line is often weak, in addition to requiring a flux $\rm S/ \rm N \geq 3$, we also only select those with line peaks at least 3$\sigma$ above the rms noise. We find that 268 galaxies meet the [\ion{O}{1}] detection criteria, of which 25 have broad [\ion{O}{1}]$\lambda6003$ component.

%%%%%%%%%%%%%%%%%%%%%%%%%%%%%%%%%%%%%%%%%%%%%%%%%%%%%%%%%%%%%%%%%%%%%%%%%%%%%%%%%%%%%%%%%%%%%%%%%%%%%%%%%%%%%%%%%%%%%%%%%%
%%%%%%%%%%%%%%%%%%%%%%%%%%%%%%%%%%%%%%%%%%%%%%%%%%%%%%%%%%%%%%%%%%%%%%%%%%%%%%%%%%%%%%%%%%%%%%%%%%%%%%%%%%%%%%%%%%%%%%%%%%

\section{Properties of the Outflow Candidates}
\label{sec:outflow_candidates}
We identify 398 galaxies with reliable ionized outflow signatures in their [\ion{O}{3}]$\lambda5007$ line, which is $\sim$1\% of our parent sample. In this section, we classify galaxies with outflows using narrow-line diagnostic diagrams, estimate BH masses for objects with detectable broad H$\alpha$ emission, and determine the properties of both the outflows and their host galaxies. {Additionally, we present findings on outflows in low-mass galaxies}.

%%%%%%%%%%%%%%%%%%%%%%%%%%%%%%%%%%%%%%%%%%%%%%%%%%%%%%%%%%%%%%%%%%%%%%%%%%%%%%%%%%%%%%%%%%%%%%%%%%%%%%%%%%%%%%%%%%%%%%%%%%

\subsection{Narrow-line Diagnostic Diagrams}
\label{sec:narrow_diag}

The harder radiation field of AGNs can result in higher fluxes of forbidden lines such as [\ion{N}{2}]$\lambda6583$, [\ion{S}{2}]$\lambda\lambda 6716, 6731$, and [\ion{O}{1}]$\lambda 6003$ lines with respect to the Balmer lines. This enables us to separate AGNs and SF galaxies when these line ratios are plotted in two-dimensional narrow-line diagnostic diagrams. 

\begin{figure*}[tbph]
\centering
\includegraphics[width=\textwidth]{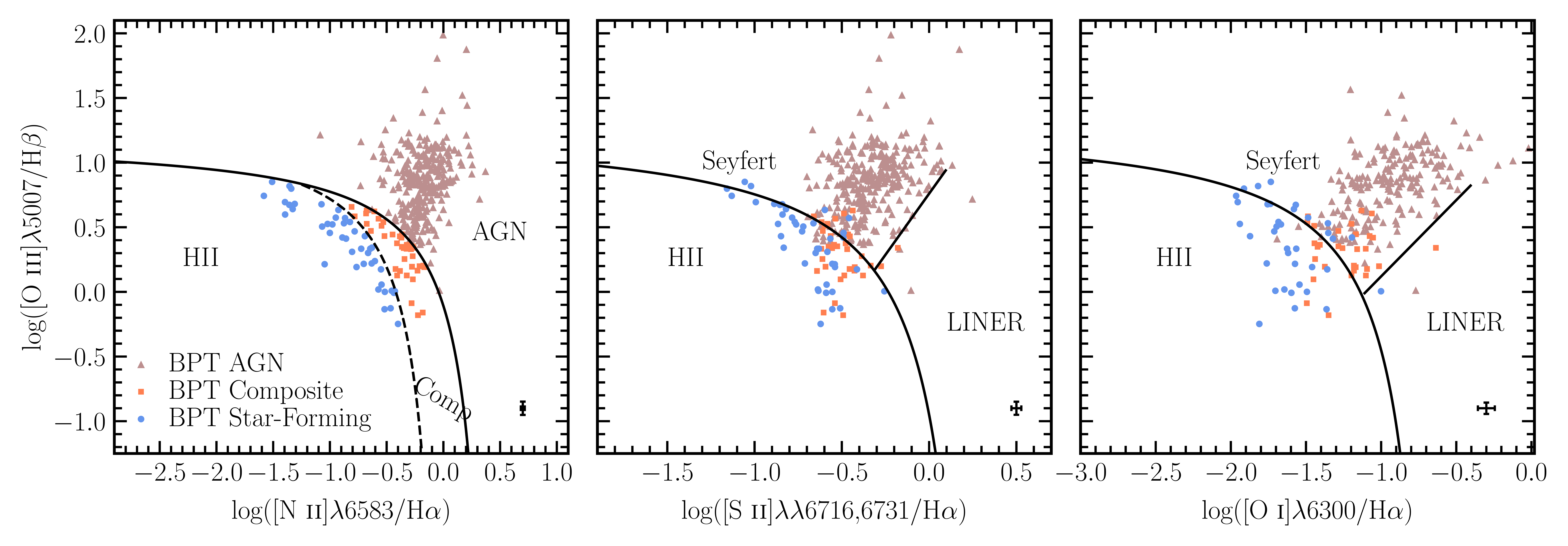}
\caption{The left panel shows the [\ion{O}{3}]/H$\beta$ vs.\ [\ion{N}{2}]/H$\alpha$ narrow-line diagnostic diagram for our outflow candidates in the GAMA survey using the classification scheme summarized in \citet{Kewley:2006}. Here 394/398 of the outflow galaxies with reliable emission lines relevant to this diagram are plotted, of which 312 are AGNs (rosy brown triangles), 39 are composites (coral squares), and 43 are SF galaxies (cornflower blue circles). The middle and right panels show these objects in the [\ion{O}{3}]/H$\beta$ vs. [\ion{S}{2}]/H$\alpha$ and [\ion{O}{1}]/H$\alpha$ diagrams. Only galaxies with reliable emission lines are plotted in these panels. Characteristic error bars are located in the lower right region of each panel. See Section~\ref{sec:narrow_diag} for details. 
}
\label{fig:bpt}
\end{figure*}

We employ the [\ion{O}{3}]/H$\beta$ vs.\ [\ion{N}{2}]/H$\alpha$ Baldwin-Phillips-Terlevich (BPT) diagnostic diagram \citep{Baldwin:1981} to classify our outflow galaxies as shown in the left panel of Figure \ref{fig:bpt}. In this diagram, the AGNs fall above the maximum starburst line from \citet{Kewley:2006}, while the SF objects occupy the area below the composite line from \citet{Kauffmann:2003}. Composite objects that fall between the two lines have both contributions from AGN and SF activity.

We only include objects with reliable emission line measurements relevant to this diagram, which is 394/398 of the outflow galaxies. {We classify 79\% of these objects as AGNs, 10\% as composites, and 11\% as SF-dominated galaxies.} Therefore, the vast majority of our galaxies with outflow signatures also have AGN photoionization signatures in their spectra, which is in agreement with previous studies that found a higher incidence of outflows in AGNs than SF galaxies \citep[e.g.,][]{Concas:2017,Matzko:2022}. This result, in addition to the large outflow velocities we find for the AGNs/composites (see Sections \ref{sec:results_outflow_vel} and \ref{sec:discussion}), suggests that these outflows are predominantly driven by AGN feedback. {For simplicity, we will refer to the AGNs and composites collectively as AGNs throughout the rest of the paper.}

%%%%%%%%%%%%%%%%%%%%%%%%%%%%%%%%%%%%%%%%%%%%%%%%%%%%
%%%%%%%%%%%%%%%%%%%%%%%%%%%%%%%%%%%%%%%%%%%%%%%%%%%%

We also plot the outflow galaxies in the [\ion{O}{3}]/H$\beta$ vs.\ [\ion{S}{2}]/H$\alpha$ and [\ion{O}{1}]/H$\alpha$ diagrams \citep{Veilleux:1987} as shown in the middle and right panels of Figure \ref{fig:bpt}. In these plots, the theoretical extreme starburst line from \citet{Kewley:2001} separates the AGNs from SF galaxies, {while the Seyfert/LINER line differentiates the Seyfert and low ionization nuclear emission region (LINER) objects. Among the BPT AGNs, $\sim$86\% with reliable [\ion{S}{2}] emission are located in the AGN region of the [\ion{S}{2}]/H$\alpha$ diagram. In the [\ion{O}{1}]/H$\alpha$ diagram, 62\% of those with detected [\ion{O}{1}] are classified as AGNs. Additionally, we find that 14\% and 26\% of the BPT-SF objects appear AGN-like in the [\ion{S}{2}]/H$\alpha$ and [\ion{O}{1}]/H$\alpha$ diagrams, respectively, while 9\% are AGNs in both diagrams.}

The small number of LINER galaxies in our outflow sample (7 in the [\ion{S}{2}]/H$\alpha$ and 7 the in [\ion{O}{1}]/H$\alpha$ diagrams) is in agreement with what \citet{Matzko:2022} found and in contrast with \citet{Hermosa:2022}, where they found outflow signature in 50\% of their LINER candidates. We note that \citet{Hermosa:2022} employ Integral Field Unit (IFU) data in their study which has a higher resolution than our sample and that of \citet{Matzko:2022}.

\subsection{Broad H$\alpha$ and BH Mass Estimates}\label{sec:broad_HA}
%%%%%%%%%%%%%%%%%%%%%%%%%%%%%%%%%%%%%%%%%%%%%%%%%%%%
\begin{figure*}[tbph]
\centering
\includegraphics[width=\textwidth]{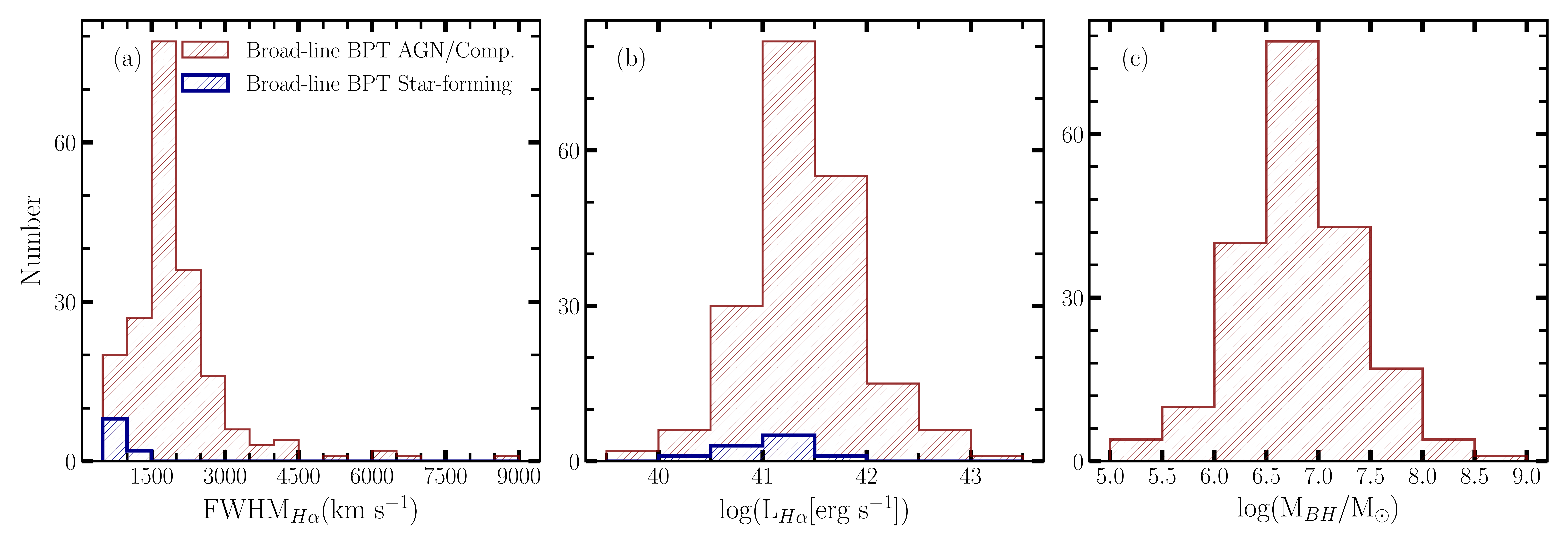}
\caption{Distributions of Broad H$\alpha$ emission parameters {and BH mass}. {Panels (a) and (b) display the histograms of the FWHM and log luminosity of the broad H$\alpha$ component for AGNs shown in maroon, and SF galaxies shown in blue. 
Virial BH mass distribution for the broad-line {AGNs} is plotted in panel (c). We estimate these BH masses using Equation (5) from \citet{Reines:2013}. Broad H$\alpha$ emissions in star-forming objects are not considered strong indicators of AGNs, and we do not estimate virial BH masses for these galaxies. For further details, refer to Section \ref{sec:broad_HA}.}}
\label{fig:br_disp}
\end{figure*}
%%%%%%%%%%%%%%%%%%%%%%%%%%%%%%%%%%%%%%%%%%%%%%%%%%%%
The presence of a broad H$\alpha$ line can be associated with rotating gas in the BLR around a BH. {By utilizing the measured parameters of this line and employing standard virial techniques, we can estimate the mass of the central BH using the formula $\rm M_{BH} \propto R_{BLR}\Delta V^2/G$. Here} gas velocity comes from the broad line width and the radius of the BLR is known to scale with the broad line luminosity based on reverberation-mapped AGNs. However, in SF galaxies, broad H$\alpha$ emission can be due to stellar activities such as supernovae, which can be transient and fade over time \citep[e.g.,][]{Baldassare:2016}. Given this, we do not accept the broad H$\alpha$ component in SF galaxies as a solid AGN indicator and do not estimate their BH mass.

As described in Section \ref{sec:analysis_other_lines}, we identify broad H$\alpha$ emission in 206 galaxies with outflow signatures. {The majority of these objects are classified as AGNs, with only 10 found among SF galaxies. The FWHM of the broad H$\alpha$ lines varies from 501 to 8607 km s$^{-1}$. The median FWHM for AGNs is 1789 km s$^{-1}$, while for SF galaxies, it is 734 km s$^{-1}$. AGNs also display greater luminosities that scale from 10$^{39.7-43.1}$ erg s$^{-1}$, with a median of 10$^{41.4}$ erg s$^{-1}$, compared to the range of 10$^{40.2-41.8}$ erg s$^{-1}$ and a median of 10$^{41.2}$ erg s$^{-1}$ in the SF objects}. We summarize these quantities for the AGNs in Table \ref{tab:bh_masses} and show their distributions in Figure \ref{fig:br_disp}.

{We estimate the virial BH masses for broad-line AGNs using the following equation \citep{Reines:2013}:}

\begin{align}\label{eq:bh_mass}
    \mathrm{log}(\frac{M_{BH}}{M_\odot}) = &\  \mathrm{log \epsilon} + \mathrm{6.57} + \mathrm{0.47log}(\frac{L_{H\alpha}}{10^{42}\ \mathrm{erg\ s^{-1}}}) \nonumber \\ 
    & +\mathrm{2.06 log(\frac{FWHM_{H\alpha}}{10^3\ km\ s^{-1}}})
\end{align}
{where $\epsilon\ = \ 1$ and $L_{H\alpha}$ represents the luminosity of the broad H$\alpha$ line.}
{The BH masses we calculate range from 10$^{5}$ to 10$^{8.6}$ M$_\odot$, with a median value of 10$^{6.8}$ M$_\odot$. These values are listed for each object in Table \ref{tab:bh_masses} and their distribution is shown in the last panel of Figure \ref{fig:br_disp}}.  We note that these BH masses are derived assuming negligible outflowing material in the BLR region. However, with the presence of outflows in this region, the reliability of this method has been debated \citep[e.g.,][]{Collin:2006,Vestergaard:2006} and alternative methods have been proposed \citep[e.g.,][]{Murray:1995,Everett:2005,Proga:2007}.

%\startlongtable
\begin{deluxetable}{ccccc}
\tabletypesize{\scriptsize}
%\setlength{\tabcolsep}{2pt}
%\tablewidth{1pt}
\tablecaption{Broad-line AGNs}
\tablehead{
\colhead{CATAID}&\colhead{{BPT Classification}}&\colhead{log $L(H\alpha)_b$} & \colhead{FWHM(H$\alpha$)$_b$}&\colhead{log $M_{BH}$}
}
\decimalcolnumbers
\startdata
1889137 &   Comp. & 41.84 &     1474 &   6.8 \\
1069351 &     AGN & 41.58 &     1684 &   6.8 \\
2132672 &     AGN & 41.64 &     2281 &   7.1 \\
1785686 &     AGN & 42.23 &     2808 &   7.6 \\
1890557 &     AGN & 41.77 &     1774 &   7.0 \\
1896259 &     AGN & 41.55 &     1589 &   6.8 \\
2379296 &     AGN & 41.22 &     1952 &   6.8 \\
1485133 &     AGN & 41.48 &     1841 &   6.9 \\
1982957 &     AGN & 41.92 &     4065 &   7.8 \\
1819774 &     AGN & 41.19 &     1371 &   6.5 \\
\enddata
\tablecomments{Column 1: Unique ID of the GAMA object. Column 2: Classification of the object in the BPT diagram. Column 3: The luminosity of the broad H$\alpha$ component in units of erg s$^{-1}$. Column 4: The width (FWHM) of the broad H$\alpha$ component in units of km s$^{-1}$, corrected for instrumental resolution. Column 5: The virial BH mass in units of M$_\odot$ by assuming the broad H$\alpha$ emission is associated with the BLR. We only include BPT-AGNs and composites in this table. See section \ref{sec:broad_HA} for more details. The entirety of Table~\ref{tab:bh_masses} is published in the electronic edition of {\it The Astrophysical Journal}. We show a portion here to give information on its form and content.}
\label{tab:bh_masses}
\end{deluxetable}

%%%%%%%%%%%%%%%%%%%%%%%%%%%%%%%%%%%%%%%%%%%%%%%%%%%%%%%%%%%%%%%%%%%%%%%%%%%%%%%%%%%%%%%%%%%%%%%%%%%%%%%%%%%%%%%%%%%%%%%%%%
\subsection{Outflow Properties}\label{sec:outflow_properties}

\subsubsection{Outflow Velocity}\label{sec:results_outflow_vel}
\begin{figure*}[tbph]
\centering
\includegraphics[width=\textwidth]{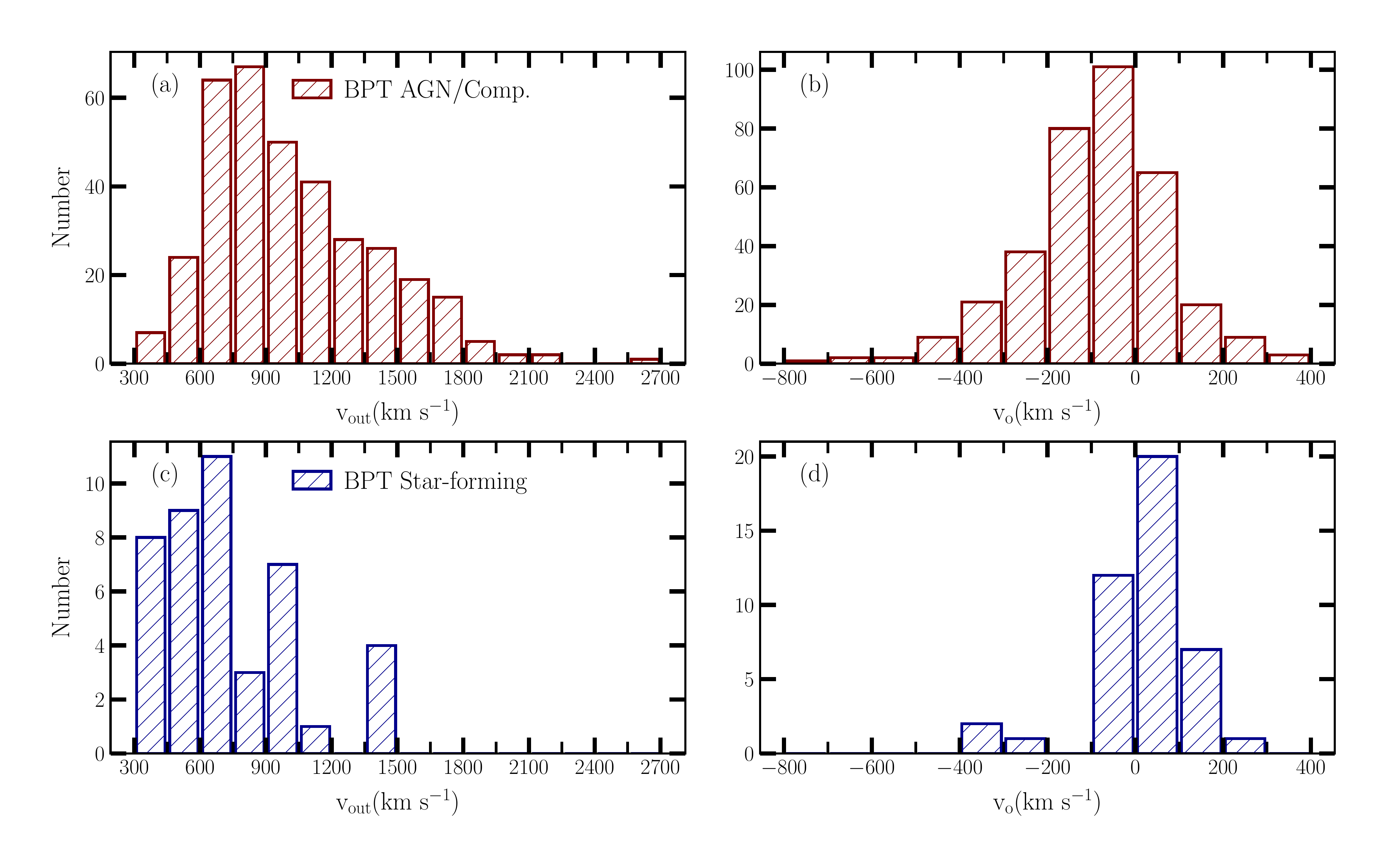}
\caption{Outflow properties. \textit{(a)--(b)}: Distributions of outflow velocity ($\rm v_{out}$) and offset velocity ($\rm v_o$) for the BPT-AGNs and composites in our outflow sample shown in maroon histograms.  
The medians of outflow velocity and offset velocity are 936 km s$^{-1}$ and $-84$ km s$^{-1}$, respectively.
\textit{(c)--(d)}: Same as panels (a)--(b) except for the SF galaxies with outflow signatures plotted in blue histograms.
Outflow and offset velocity median values are 655 km s$^{-1}$ and 28 km s$^{-1}$, respectively.  See Section \ref{sec:outflow_properties} for more details. 
}
\label{fig:vel}
\end{figure*}

{As discussed in Section \ref{sec:analysis_outflow_vel}, W$_{80}$ measures the outflow velocity in our sample, which varies from 327 to 2689 km s$^{-1}$. The AGNs exhibit higher outflow velocities, with a median of 936 km s$^{-1}$, while SF galaxies are found with a median of 655 km s$^{-1}$. For the AGN candidates with broad H$\alpha$ detection, the median outflow velocity is slightly higher at 961 km s$^{-1}$, compared to 880 km s$^{-1}$ for those without broad H$\alpha$ lines. Histograms of these velocities {for each galaxy classification} are shown in panels (a) and (c) of Figure \ref{fig:vel}. Additionally, the outflow velocity for each object is detailed in Table \ref{tab:gal_prop}.}

{Outflows in AGNs and SF galaxies have been observed with velocities ranging from a few hundred to thousands of km s$^{-1}$ \citep[e.g.,][]{Harrison:2014,Mullaney:2013,Matzko:2022,Aravindan:2023}. Some studies suggest that velocities of at least 500 km/s are associated with AGN activity \citep{Mullaney:2013}, as these require an outflow power that exceeds that of starbursts \citep{Fabian:2012}. This is also related to the maximum linewidths observed in galaxy dynamics and mergers in high-redshift ultraluminous infrared galaxies \citep[][and the references within]{Harrison:2012}. We find that $\sim$97\% of the {AGNs} have outflow velocities of $\rm W_{80}>500$ km s$^{-1}$, indicating that AGN-driven mechanisms are responsible for the outflows in these galaxies.}

Our results are consistent with the literature.
The median outflow velocities in the starburst galaxies in \citet{Rupke:2002,Rupke:2005} were $\sim 300$ km s$^{-1}$, while \citet{Hill:2014} and \citet{Rupke:2013} found median velocities of order $\sim 600$ km s$^{-1}$ for their starburst objects. \citet{Matzko:2022} found outflow velocities on the order 700 km s$^{-1}$ for their AGNs and lower velocities with an average of 300 km s$^{-1}$ for their SF galaxies. \citet{Harrison:2012} found a median bulk outflow velocity of 780 km s$^{-1}$  for their type 2 AGN sample while \citet{Mullaney:2013} found a mean of $~900$ km s$^{-1}$ for their type 1 AGNs. \citet{Zakamska:2014} found a median velocity of 752 km s$^{-1}$ for their luminous obscured quasars.

\subsubsection{Offset Velocity}\label{sec:offset_vel}
{The offset velocity v$_0$  in our sample, which is the separation between the [\ion{O}{3}] line components, ranges from $-779$ to 386 km s$^{-1}$. The {AGNs} show a blueshifted median $\rm v_o$ of $-84$ km s$^{-1}$, while the SF objects exhibit a redshifted median of 28 km s$^{-1}$. We find that 32\% of the outflows are redshifted (v$_0>0$), where the incidence of redshifted lines in {AGNs} is 28\%, compared to 65\% in SF galaxies.  Furthermore, 27\% of {AGNs} with broad H$\alpha$ detection are found with redshifted second components, whereas 29\% of AGNs without broad H$\alpha$ have this feature. {Histograms of the $\rm v_o$ for the AGNs and SF galaxies are shown in panels (b) and (d) of Figure \ref{fig:vel}, and Table \ref{tab:gal_prop} includes this velocity for each object.  
}
}

{The offset velocity is sensitive to dust extinction \citep{Bae:2016}, and in the presence of an obscured central region (like an AGN), the blueshifted broader component can trace outflows in the NLR around a BH. Hence, such lines are often attributed to an AGN origin
\citep{Harrison:2012,Liu:2020,Matzko:2022}. On the other hand, the redshifted outflow lines can be a consequence of the orientation of the galaxy to the line of sight \citep{Bae:2016}, and have also been reported in AGNs \cite[e.g.,][]{Crenshaw:2010,Mullaney:2013}. Therefore, both the redshifted and blueshifted outflow components among our {AGNs} with velocities over 500 km s$^{-1}$ likely signify AGN feedback. 
In contrast, stellar-driven outflows do not necessarily originate from the center of galaxies and can occur at any location, thus they generally are not affected by extinction \citep{Aravindan:2023} and are typically observed with symmetric outflow components \citep[e.g.,][]{Concas:2017,Davies:2019,Matzko:2022}.}

These results are consistent with previous studies that suggest mostly broad blueshifted outflow lines in AGNs and a more symmetric broad line in SF galaxies \citep[e.g,][]{Harrison:2014,Concas:2017,Manzano:2019,Matzko:2022,Aravindan:2023}. 
The ratio of our redshifted outflow lines is similar to the incidence ratio of 28\% in \citet{Barth:2008}, while it is larger than the 6\% in \citet{Greene:2005} and \citet{Crenshaw:2010}.

%%%%%%%%%%%%%%%%%%%%%%%%%%%%%%%%%%%%%%%%%%%%%%%%%%%%%%%%%%%%%%%%%%%%%%%%%%%%%%%%%%%%%%%%%%%%%%%%%%%%%%%%%%%%%%%%%%%%%%%%%%

\subsection{Host Galaxy Properties}\label{sec:host_prop}
\begin{figure*}[tbph]
\centering
\includegraphics[width=\textwidth]{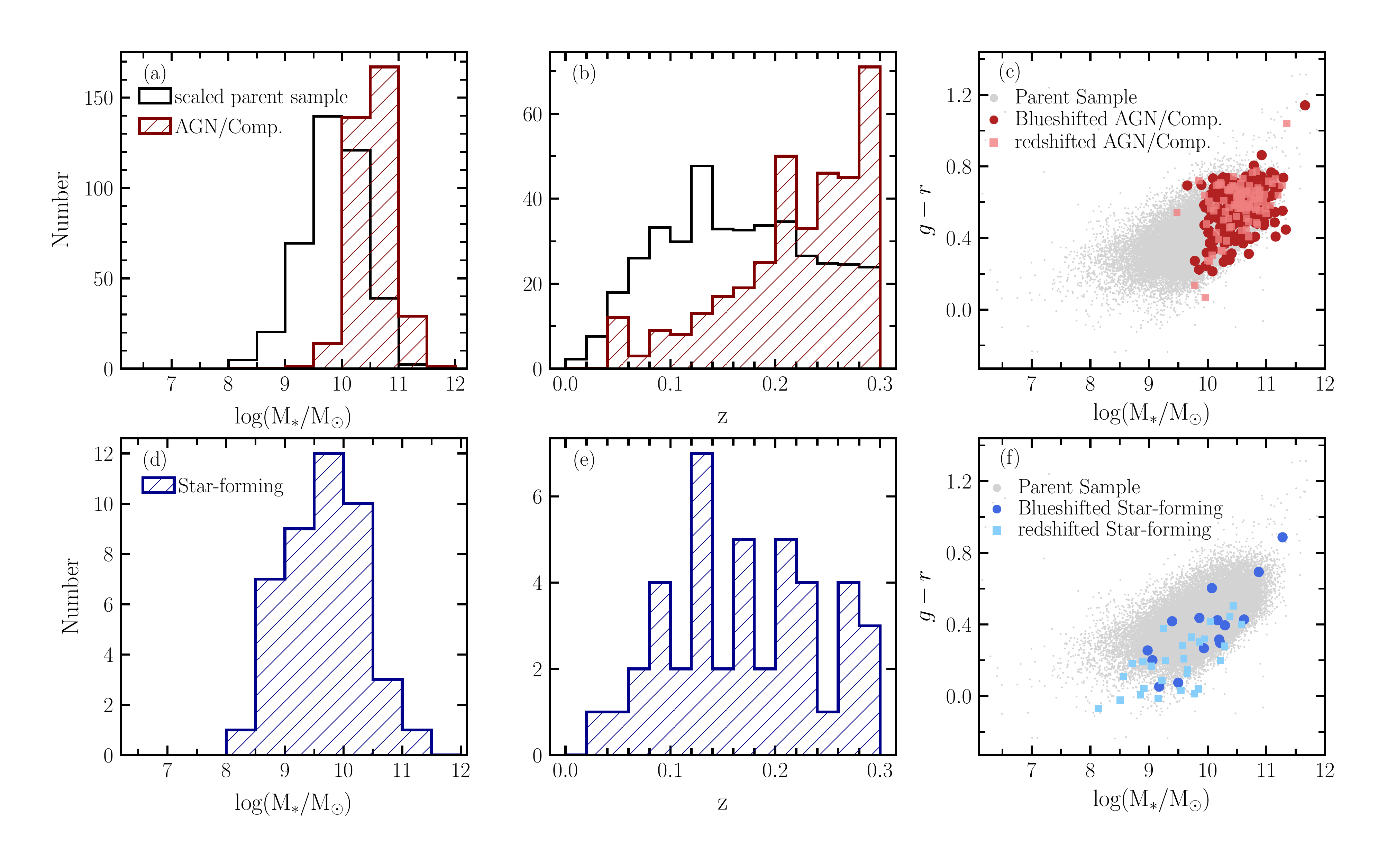}
\caption{Host galaxy properties for the outflow candidates. \textit{(a)--(c)}: Panels (a) and (b) show the distributions of host galaxy stellar mass and redshift (hashed maroon histograms) for AGNs/composites. Our parent sample (normalized to the number of outflow galaxies) is also shown as a black histogram. Panel (c) shows the $g-r$ vs.\ log(M$_*/$M$_\odot$) plot for AGNs. Here the distribution of the blueshifted and redshifted outflow components are displayed as red circles and pink squares, respectively.  All values are adopted from \texttt{StellarMassesG02CFHTLS v24} and \texttt{StellarMassesGKV v24} tables \citep{Taylor:2011,Bellstedt:2020}. No noticeable difference between the host properties of the blueshifted and redshifted outflow lines can be seen. 
\textit{(d)--(f)}: Same as panels (a)--(c) except for the SF galaxies with outflow signatures (hashed blue histograms). In panel $f$, the blueshifted and redshifted lines are shown as dark blue circles and light blue squares, respectively. See Section \ref{sec:host_prop} for more details. }
\label{fig:gal_prop}
\end{figure*}

We plot the stellar mass distribution of outflow hosts separated by their classifications in panels (a) and (d) of Figure \ref{fig:gal_prop}. The lowest host galaxy mass belongs to an SF object with a stellar mass of $10^{8.1}$ M$_\odot$, while the highest mass is a BPT-AGN with a stellar mass of $10^{11.7}$ M$_\odot$. The median log galaxy mass for the SF galaxies and AGNs are 9.7 and 10.6 M$_\odot$, respectively. $\sim96$\% of the AGNs have stellar masses M$_*>10^{10}$ M$_\odot$, while 33\% of the SF galaxies are within this mass range. 

%%%%%%%%%%%%%%%%%%%%%%%%%%%%%%%%%%%%%%%%%%%%%%%%%%%%
%%%%%%%%%%%%%%%%%%%%%%%%%%%%%%%%%%%%%%%%%%%%%%%%%%%%

The redshift distribution of AGNs and SF galaxies can be found in panels (b) and (e) of Figure \ref{fig:gal_prop}, respectively. We select objects with redshift $\rm z<$0.3 by design to ensure the [\ion{S}{2}] doublet is covered in the observed spectra. The median redshift for our AGNs and SF galaxies are  0.23 and 0.17, respectively.

We show the color-mass diagram for the AGNs in panel (c) of Figure \ref{fig:gal_prop}. These objects are predominantly among more massive galaxies and follow a similar color range as the parent sample in the high mass range. There does not appear to be a significant difference between the AGN host galaxies with redshifted and blueshifted outflow lines. However, the median color of the galaxies with redshifted lines is 0.6 compared to the slightly bluer median color of 0.57 for the host galaxies with blueshifted lines.

The color-mass diagram for the SF galaxies is plotted in panel (f) of Figure \ref{fig:gal_prop}. Here, the redshifted lines in the SF galaxies seem to be among less massive and bluer objects. The bluer color may indicate that the star formation has not been impacted by outflows \citep{Aravindan:2023}. In contrast, the SF galaxies with blueshifted broad lines also extend to redder and higher masses.

%%%%%%%%%%%%%%%%%%%%%%%%%%%%%%%%%%%%%%%%%%%%%%%%%%%%%%%%%%%%%%%%%%%%%%%%%%%%%%%%%%%%%%%%%%%%%%%%%%%%%%%%%%%%%%%%%%%%%%%%%%
\subsection{Low-Mass Galaxies}\label{sec:low_mass}
While stellar feedback has been considered the main source of feedback in low-mass galaxies, theoretical models have attained contrasting results on the extent of AGN feedback and the impact of SF-driven outflows in them \citep{Angl:2017,Trebitsch:2018,Dashyan:2018,Koudmani:2019,Koudmani:2021,Barai:2019,Sharma:2020}. Given the recent observations of AGNs in low-mass/dwarf galaxies \citep[e.g.,][]{Reines:2013,Moran:2014,Molinafex:2021,Salehirad:2022} as well as the evidence of outflows in this mass range \citep{Liu:2020,Aravindan:2023}, it is important to search for AGN feedback specifically in the low-mass regime.

In this work, we identify 45 galaxies with masses M$_*<10^{10}$ M$_\odot$ among the outflow candidates, of which 11 are BPT AGNs, 4 are composites, and 29 are classified as SF galaxies. The remaining low-mass galaxy has an unreliable H$\alpha$ measurement and thus is not classified. All the AGNs/composites are among the \citet{Salehirad:2022} sample. 
We have found that while AGN hosts are more common in our overall outflow sample, star-forming galaxies are the primary hosts of outflows in our low-mass galaxies. However, selection effects could contribute to this finding. For example, the BPT diagram has difficulty distinguishing AGNs in the low-mass range, and low-metallicity AGNs can overlap with low-metallicity starbursts in this diagram \citep{Groves:2006}.

\begin{figure*}[tbph]
\centering
\includegraphics[width=\textwidth]{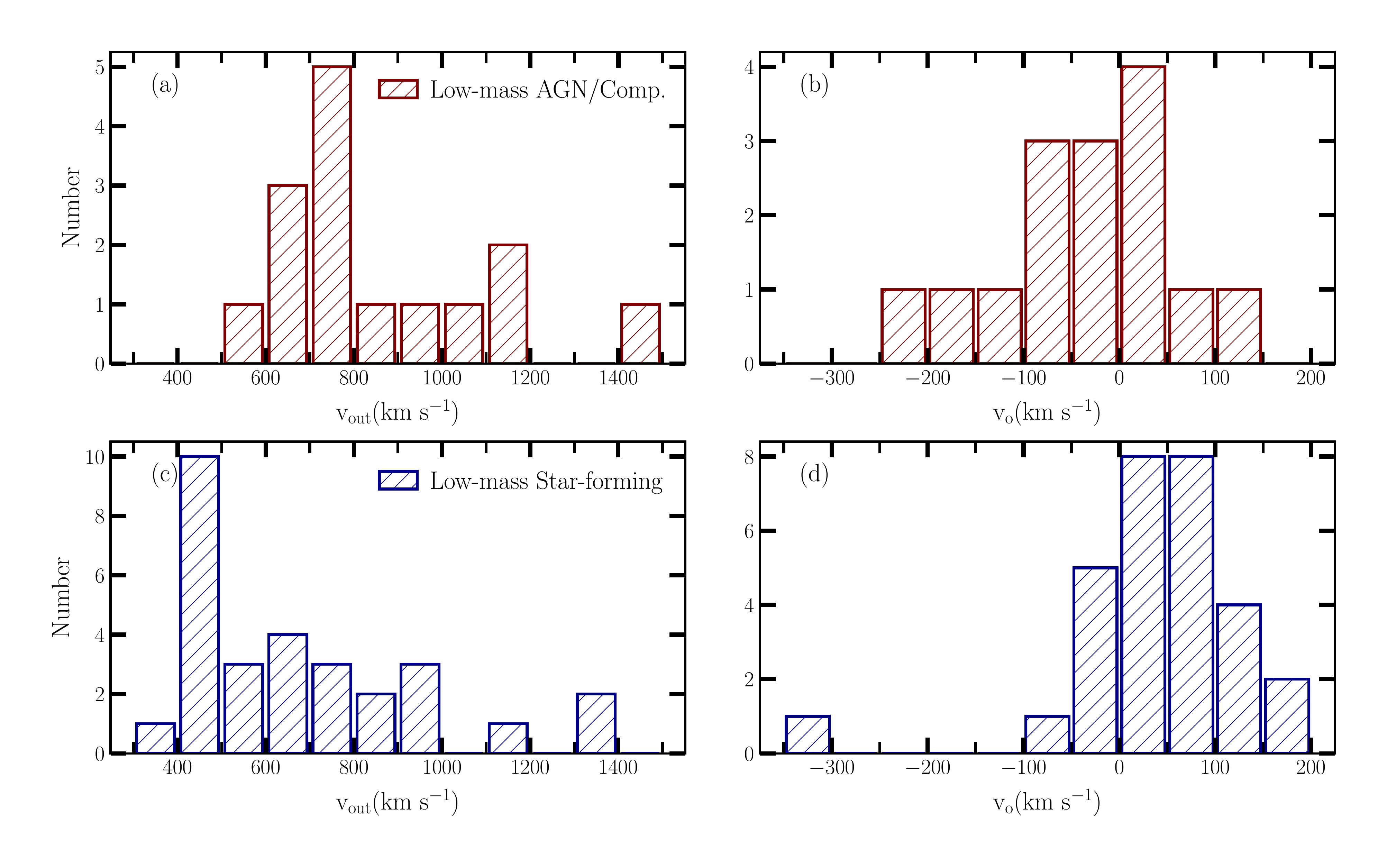}
\caption{Same as Figure \ref{fig:vel}, but for the low-mass outflow galaxies with masses $\rm M_*<10^{10}\ \rm M_\odot$. 
The medians of outflow and offset velocities for AGNs and composites are 777 and $-46$ km s$^{-1}$, respectively, while SF galaxies have a lower median outflow velocity of 609 km s$^{-1}$ and a redshifted median offset velocity of 42 km s$^{-1}$. See Section \ref{sec:low_mass} for more details.
}
\label{fig:vel_dwarfs}
\end{figure*}

The outflow velocities among the low-mass galaxies span a range of 327 to 1449 km s$^{-1}$, where the median outflow velocity for the AGNs and SF galaxies are 777 km s$^{-1}$ and 609 km s$^{-1}$, respectively. Distributions of outflow velocities are shown in panels (a) and (c) of Figure \ref{fig:vel_dwarfs}. {Our median values are higher than what \citet{Aravindan:2023} and \citet{Liu:2020} found for their SF and AGN samples (484 km s$^{-1}$) and less than the weighted averages reported in \citet{Manzano:2019}.}

{The offset velocities vary from $-303$ to 191  km s$^{-1}$, with a median of $-46$ km s$^{-1}$ for AGNs and a median of 42 km s$^{-1}$ for SF galaxies. One of the SF objects is found with a blueshifted velocity of $\sim -300$ km s$^{-1}$, where the [\ion{O}{3}] line components have similar widths and resemble the line profile shown in panel (d) of Figure \ref{fig:red_dp_lines_examples}. }
We display the distributions of offset velocities for low-mass galaxies in panels (b) and (d) of Figure \ref{fig:vel_dwarfs}. {Previous studies such as \citet{Manzano:2019} found an offset velocity of -108  km s$^{-1}$ for their AGN candidates and 5 km s$^{-1}$ for their SF galaxies. }
\citet{Liu:2020} found the average offset velocity of $-64$ km s$^{-1}$ for their AGN sample, while  \citet{Aravindan:2023} found an average offset velocity of 0 km s$^{-1}$ for their SF galaxies, which was $-60$ km s$^{-1}$ if only blueshifted regions were considered. If we include only the blueshifted outflow lines, our medians for AGNs and SF galaxies change to $-62$ and $-34$ km s$^{-1}$, respectively. 

Outflow signatures among our low-mass AGNs suggest that BH feedback exists in this mass range and should be considered as a factor in the evolution models of low-mass galaxies. Moreover, we find that outflows in AGNs are typically faster and blueshifted compared to SF galaxies, which can deplete ISM from gas and quench SF \citep{Bradford:2018} or trigger SF \citep{Schutte:2022}, again suggesting the importance of AGN feedback in low-mass galaxies.

%%%%%%%%%%%%%%%%%%%%%%%%%%%%%%%%%%%%%%%%%%%%%%%%%%%%%%%%%%%%%%%%%%%%%%%%%%%%%%%%%%%%%%
%%%%%%%%%%%%%%%%%%%%%%%%%%%%%%%%%%%%%%%%%%%%%%%%%%%%%%%%%%%%%%%%%%%%%%%%%%%%%%%%%%%%%%%%
%%%%%%%%%%%%%%%%%%%%%%%%%%%%%%%%%%%%%%%%%%%%%%%%%%%%%%%%%%%%%%%%%%%%%%%%%%%%%%%%%%%%%%
%%%%%%%%%%%%%%%%%%%%%%%%%%%%%%%%%%%%%%%%%%%%%%%%%%%%%%%%%%%%%%%%%%%%%%%%%%%%%%%%%%%%%%%%%%%%%%%%%%%%%%%%%%%%%%%%%%%%%%%%%%
%%%%%%%%%%%%%%%%%%%%%%%%%%%%%%%%%%%%%%%%%%%%%%%%%%%%%%%%%%%%%%%%%%%%%%%%%%%%%%%%%%%%%%%%%%%%%%%%%%%%%%%%%%%%%%%%%%%%%%%%%%
\section{{Summary and Conclusions}}\label{sec:discussion}

In this work, we systematically search for ionized outflow signatures in the [\ion{O}{3}]$\lambda\lambda 4959, 5007$ doublet emission line. Our parent sample consists of 39,612 galaxies with redshift $\rm z<0.3$ from the GAMA survey. We identify double-component features that may signify outflows in 398 galaxies, {of which 45 are among low-mass galaxies with stellar masses $\rm M_* <10^{10}$ $M_\odot$}. Only 8 of our outflow candidates have SDSS spectra, and thus we are presenting novel outflow candidates in this work.

We classify our outflow galaxies using the BPT diagram as shown in the left panel of Figure \ref{fig:bpt}. Of the 394 galaxies with reliable measurements of the emission lines used in this diagram, 79\% are AGNs, 10\% are composites, and the remaining 11\% are SF-dominated galaxies. Thus, the majority of our outflow sample is among galaxies with at least some level of AGN activity, which is consistent with previous work that finds a higher incidence of outflows in AGNs \citep[e.g.,][]{Concas:2017}. %reduced reference number here 

{We also search for broad H$\alpha$ emission and identify 206 galaxies, of which 196 are among AGNs/composites and 10 are SF galaxies. We estimate virial BH masses for the AGNs and composites using the broad H$\alpha$ line parameters, ranging from $10^5$ to $10^{8.6}$ $\rm M_\odot$. Distributions of the broad H$\alpha$ parameters and BH masses are shown in Figure \ref{fig:br_disp}.}

{We identify outflow components that are symmetric, blueshifted, and redshifted with respect to the systemic components of the [\ion{O}{3}] lines. In $\sim 9\%$ of the outflow candidates, either two peaks are visible by eye or they have components with similar widths (see Figure \ref{fig:red_dp_lines_examples} and Section \ref{sec:analysis_o3_lines}).
These lines can represent double-peaked lines, which can be produced by the disk rotation of the NLR around a single BH, biconical outflows, or distinct NLRs in merging AGNs \cite[e.g.,][]{Shen:2011}.} 

{We adopt W$_{80}$ to measure the outflow velocities as shown in Figure \ref{fig:w80} and find generally faster outflow velocities in BPT-AGNs and composites} with a median velocity of 936 km s$^{-1}$ compared to 655 km s$^{-1}$ in the SF galaxies. Moreover, the majority of the AGNs/composites have outflow velocities  $\rm W_{80}>500$ km s$^{-1}$, which is considered the limit that AGN feedback due to outflows is noteworthy \citep{Fabian:2012}, and indicates AGN-driven mechanisms for the outflows in these galaxies.

The offset velocity between the systemic and outflow components of the [\ion{O}{3}]$\lambda5007$ line varies from $\sim -780$ to 390 km s$^{-1}$ (see Section \ref{sec:offset_vel}). The outflows in AGNs and composites have a blueshifted median offset velocity of $-84$ km s$^{-1}$, while the SF objects have a redshifted median of 28 km s$^{-1}$. The incidence of redshifted outflows in our sample is 32\%, where the incidence among AGNs/composites and SF-dominated galaxies are 28\% and 65\%, respectively.

Host galaxy properties for our outflow sample are listed in Table \ref{tab:gal_prop} and their distributions are shown in Figure \ref{fig:gal_prop}. The host galaxy stellar masses of our outflow sample range from $10^{8.1}$ to $10^{11.7}$ $\rm M_\odot$, where the median galaxy mass of the AGNs/composites and SF galaxies are $10^{10.6}$ and $10^{9.7}$ M$_\odot$, respectively. 96\% of the AGNs have stellar masses $\rm M_*>10^{10}$, while only 33\% of the SF galaxies are within this mass range. 
The BPT AGNs and composites are predominantly among massive galaxies and follow a similar color range to our parent sample, while the SF objects are among lower-mass and bluer objects. 

Of the 45 low-mass galaxies that exhibit outflow signatures, 11 are classified as AGNs, 4 are composites, and 29 are SF galaxies. Outflows in the low-mass AGNs/composites are faster and blueshifted with median outflow and offset velocities of 777 km s$^{-1}$ and $-46$ km s$^{-1}$. On the other hand, outflows in SF objects are found with a median outflow velocity of 609 km s$^{-1}$ and a redshifted median offset velocity of 42 km s$^{-1}$, see Figure \ref{fig:vel_dwarfs}. 
{The existence of faster-moving outflows in low-mass AGNs suggests that AGN feedback is noteworthy in these objects and should be considered a factor in galaxy evolution models in this mass range.}

Identifying these novel ionized gas outflows is the first step in furthering our knowledge of feedback and its impact on the evolution of the central BHs and their host galaxies. Future studies of the molecular and neutral gas outflows associated with these objects can help us understand the mechanisms involved in producing them and how they are distributed throughout galaxies. The {\it James Webb Space Telescope} could be used to trace the molecular phase of the outflows by observing the mid-infrared rotational and rovibrational transitions of H$_2$, which can be further explored as a tracer of AGN feedback \citep{Cicone:2018}. Studying the radio luminosity of these targets allows us to explore whether the mechanical energy from a radio jet is responsible for these outflows. From the X-ray spectra and the bolometric luminosity of the AGNs,  we can investigate if the energy of radiatively driven outflows by AGNs is sufficient to couple with ISM and produce them or if the radiation from stellar processes is the more likely scenario. Finally, follow-up {integral field spectroscopy} observations of these galaxies would allow us to trace the kinematics and morphology of outflows on pc to kpc scales and investigate the impact on star formation. 

%%%%%%%%%%%%%%%%%%%%%%%%%%%%%%%%%%%%%%%%%%%%%%%%%%%%%%%%%%%%%%%%%%%%%%%%%%%%%%%%%%%%%%
%%%%%%%%%%%%%%%%%%%%%%%%%%%%%%%%%%%%%%%%%%%%%%%%%%%%%%%%%%%%%%%%%%%%%%%%%%%%%%%%%%%%%%%%
%%%%%%%%%%%%%%%%%%%%%%%%%%%%%%%%%%%%%%%%%%%%%%%%%%%%%%%%%%%%%%%%%%%%%%%%%%%%%%%%%%%%%%
\acknowledgments
We thank the anonymous reviewer for their helpful comments and suggestions that greatly improved this work. 
AER acknowledges support for this work provided by NASA through EPSCoR grant number 80NSSC20M0231 and the NSF through CAREER award 2235277. The work of M.M. is supported in part through a fellowship sponsored by the Willard L. Eccles Foundation.

GAMA is a joint European-Australasian project based around a spectroscopic campaign using the Anglo-Australian Telescope. The GAMA input catalogue is based on data taken from the Sloan Digital Sky Survey and the UKIRT Infrared Deep Sky Survey. Complementary imaging of the GAMA regions is being obtained by a number of independent survey programmes including GALEX MIS, VST KiDS, VISTA VIKING, WISE, Herschel-ATLAS, GMRT and ASKAP providing UV to radio coverage. GAMA is funded by the STFC (UK), the ARC (Australia), the AAO, and the participating institutions. The GAMA website is http://www.gama-survey.org/. Based on observations made with ESO Telescopes at the La Silla Paranal Observatory under programme ID 179.A-2004. Based on observations made with ESO Telescopes at the La Silla Paranal Observatory under programme ID 177.A-3016.

%%%%%%%%%%%%%%%%%%%%%%%%%%%%%%%%%%%%%%%%%%%%%%%%%%%%%%%%%%%%%%%%%%%%%%%%%%%%%%%%%%%%%%%%%%%%%%%%%%%%%%%%%
%%%%%%%%%%%%%%%%%%%%%%%%%%%%%%%%%%%%%%%%%%%%%%%%%%%%%%%%%%%%%%%%%%%%%%%%%%%%%%%%%%%%%%%%%%%%%%%%%%%%%%%%%
%%%%%%%%%%%%%%%%%%%%%%%%%%%%%%%%%%%%%%%%%%%%%%%%%%%%%%%%%%%%%%%%%%%%%%%%%%%%%%%%%%%%%%%%%%%%%%%%%%%%%%%%%
%%%%%%%%%%%%%%%%%%%%%%%%%%%%%%%%%%%%%%%%%%%%%%%%%%%%%%%%%%%%%%%%%%%%%%%%%%%%%%%%%%%%%%%%%%%%%%%%%%%%%%%%%

\software{
Astropy \citep{astropy2013,astropy2018},
Matplotlib \citep{matplotlib},
LMFIT\citep{lmfit}, Pandas\citep{pandas:2010data}}

%%%%%%%%%%%%%%%%%%%%%%%%%%%%%%%%%%%%%%%%%%%%%%%%%%%%%%%%%%%%%%%%%%%%%%%%%%%%%%%%%%%%%%%%%%%%%%%%%%%%%%%%%
%%%%%%%%%%%%%%%%%%%%%%%%%%%%%%%%%%%%%%%%%%%%%%%%%%%%%%%%%%%%%%%%%%%%%%%%%%%%%%%%%%%%%%%%%%%%%%%%%%%%%%%%%
%%%%%%%%%%%%%%%%%%%%%%%%%%%%%%%%%%%%%%%%%%%%%%%%%%%%%%%%%%%%%%%%%%%%%%%%%%%%%%%%%%%%%%%%%%%%%%%%%%%%%%%%%
%%%%%%%%%%%%%%%%%%%%%%%%%%%%%%%%%%%%%%%%%%%%%%%%%%%%%%%%%%%%%%%%%%%%%%%%%%%%%%%%%%%%%%%%%%%%%%%%%%%%%%%%%

\clearpage
\bibliographystyle{aasjournal}
\bibliography{papers}

\end{document}